# Shape and Chirality Transitions in Off-Axis Twist Nematic Elastomer Ribbons

Yoshiki Sawa[1], Kenji Urayama[2], Toshikazu Takigawa

Department of Material Chemistry, Kyoto University, Kyoto 615-8510, Japan

Vianney Gimenez-Pinto[1], Badel L. Mbanga[*], Fangfu Ye[**],

Jonathan V. Selinger, Robin L. B. Selinger[2]

Liquid Crystal Institute, Kent State University, Kent, OH 44242

* Current address: Physics Dept., Tufts University, Medford, MA

** Current address: Georgia Institute of Technology, School of Physics, Atlanta, GA

[1] Y.S. and V. G.-P. contributed equally to this work.

[2] To whom correspondence may be addressed. E-mail: urayama@rheogate.polym.kyoto-u.ac.jp or rselinge@kent.edu

Abstract: Using both experiments and finite element simulations, we explore the shape evolution of off-axis twist nematic elastomer ribbons as a function of temperature. The elastomers are prepared by cross-linking the mesogens with planar anchoring of the director at top and bottom surfaces with a 90° left-handed twist. Shape evolution depends sensitively on the off-axis director orientation at the sample mid-plane. When the director at midplane is parallel to either the ribbon's long or short axes, ribbons form either helicoids or spirals depending on aspect ratio and temperature. If the director at midplane is more than 5° off-axis, then they form only spiral ribbons. Samples in all these geometries show a remarkable transition from right- to left-handed chiral shapes on change of temperature. Simulation studies provide insight into the mechanisms driving shape evolution and enable engineering design of these materials for future applications.

## Introduction

Liquid crystal elastomers (LCEs) possess a notable feature: their macroscopic shape and molecular orientation are strongly coupled due to the combination of liquid crystallinity and rubber elasticity. As a prominent example of this coupling effect, nematic elastomers (NEs) with uniform director alignment exhibit uniaxial deformation of more than several tens percent by a finite temperature change across the isotropic-nematic transition; the stretching and contraction are caused by an increase and decrease in nematic order resulting from cooling and heating, respectively [1][2][3]. More densely cross-linked nematic networks experience smaller strains [4]. Recently, Broer et al. [5][6] and White et al. [7] reported that several inhomogeneous types of deformation such as bending and torsion were thermally driven for highly crosslinked rigid LC networks with inhomogeneous configurations of director. They employed the "hybrid" alignment where the nematic director has a continuous spatial gradient across the film, with homeotropic alignment on one surface and planar alignment on the other. They also considered the "twist" geometry with planar anchoring on top and bottom surfaces, with the director smoothly twisting by $90^o$ in between. In both configurations, gradients in director orientation give rise to inhomogeneous distortions, driving an initially flat sample to deform into a three-dimensional structure. These studies demonstrated that controlling the geometry of the director and its spatial variation in the sample can drive various types of temperature-induced deformation in these materials including bending and twisting of the sample at the macroscopic level. These results showed promise for potential applications of LC networks as thermally-driven soft actuators [8]. Simple motors have also been demonstrated that are driven by LC networks' response to illumination [9] or changes in humidity [10].

In previous work [11], we demonstrated that NE films with 90° twist configuration (TNEs) formed different chiral ribbon morphologies in response to temperature change. Studies of TNE ribbons also provide insight into mechanisms for interaction of molecular-scale and macroscopic chirality in soft materials. Lipid assemblies [12][13], protein fibrils [14] and twist nematic elastomer ribbons can form a range of chiral shapes, including cylindrical tubes with "barber-pole" markings; saddle-like helicoids with negative Gaussian curvature, as shown in Fig. 1(a); and spiral ribbons that are hollow, with zero Gaussian curvature, as shown in Fig. 1(b) [11][15][16]. Theoretical studies [11][17][18] have examined shape selection of tubules, helicoids and spiral ribbons. In contrast to the chiral supramolecular aggregates of the order of nano- or micrometer scale, the shape formation of TNE films occurs at the macroscopic scale, with structures of the order of milli- or centimeter, and the dimensional parameters of the TNE films can be experimentally varied as desired. These features are advantageous to the characterization of the structural parameters of the shapes and to the investigation of the dimensional effect on the shape formation. Further, the TNEs, i.e., loosely cross-linked LC networks are expected to vary in shape more dramatically in response to change of nematic order than more highly cross-linked glassy materials, because the elastomeric feature of the TNEs results in a considerably stronger coupling of mesogen orientation and network deformation. It was demonstrated for a NE with hybrid alignment that the elastomeric feature boosted the degree of thermally driven bending remarkably [19].

Our previous work [11] demonstrated both experimentally and theoretically how TNE ribbons can deform into helicoids or spirals on change of temperature, depending on their aspect ratio. We examined TNE ribbons with two types of twisted microstructure, designated as L- and S-geometries: the nematic order orientation changes smoothly in the clockwise direction by 90° from the bottom to the top surface with the director at the mid-plane parallel to the long (L) or short (S) axis of the film, as shown in Figure 2. Under temperature change, narrow TNE films twisted around their central lines and formed helicoids. When the sample's width-thickness ratio exceeded a critical value, the central line of the film curved into helices and produced spiral ribbons. L- and S- geometry TNE films also displayed a fascinating ***reversal of macroscopic chirality as a function of temperature,*** with diverging chiral pitch at the transition. We determined that these shape transitions depend on three parameters: width-thickness ratio, temperature, and twist geometry.

In the present work, we investigate shape selection of TNE ribbons with an *off-axis* twist geometry (designated as X-geometry), which was not examined in the previous study [11], as a function temperature and aspect ratio. Here the director at the mid-plane is not parallel to either the long or short axis of the ribbon. The X-geometry is characterized by the angle $\theta$ formed by the director at the bottom surface and the ribbon's long axis (-45° ≤ $\theta$ ≤ +45°) as shown in Figure 2. TNE-X samples thus have a lower degree of symmetry in microstructure, offering the possibility for different types of actuation. Our numerical studies elucidate the shape and macroscopic helical/spiral chirality of TNE-X ribbons as a function of angle $\theta$. This study will contribute to a deeper understanding of complex shape evolution and macroscopic chirality selection and their dependence on microstructure in chiral soft matter.

**Experimental**

Sample preparation

The TNE specimens were prepared by the "chiral imprinting" method described elsewhere [11]. The side-chain type TNEs were made by the photopolymerization between monoacrylate mesogen (A-6OCB) and diacrylate cross-linker (HDDA) with photoinitiator (IRGACURE® 784). The cross-linker concentration in feed was 7 mol%. The nonreactive miscible nematic solvent (6OCB) was mixed with the reactants to broaden the temperature range of the nematic phase. The mixing ratio ([A-6OCB]/[6OCB]) was 5/4 by weight. The chemical structures of the compounds are shown in Figure 3. The reactant mixtures were loaded in the TN glass cells with a gap of 40 or 100 μm. The surfaces of the top and bottom substrates were coated by uniaxially rubbed polyimide layer, and the rubbing directions of the two substrates were crossed with each other. The total twist angle of the mesogen orientation between the two substrates was controlled to be 90° by adding an optimized amount of the nonreactive chiral dopant S-811 (0.060 and 0.015wt% for 40μm and 60μm-thick cells, respectively) which induced a left-handed twist configuration of the mesogens.

The TN cells were irradiated using a xenon lamp with emission at a wavelength 526 nm for 30 min. The temperature for the photopolymerization was 318 K in the nematic state. After the reaction, the cells were immersed in dichloromethane for a few days until the gel film detached from the substrates due to the swelling pressure. The unreacted and nonreactive materials including the chiral dopant were washed out from the detached gel films by renewing dichloromethane several times. The gel films were gradually de-swollen by adding methanol (poor solvent) stepwise, and the fully de-swollen films were dried in air. The dried TNE films considerably curled at room temperature due to a large volume reduction of ca. 50% from the preparation state and a finite temperature difference from the preparation temperature. The ribbon specimens were cut out from the large sheet films at ca. 350 K where the sheet became flat, so that the orientation of the bottom surface could have an angle ($\theta$) relative to the long axis of the ribbon (Figure 2). The angle $\theta$ was determined by the polarizing optical microscopy with an accuracy of ±1°. We employed the two TNE ribbon specimens with almost the same length, width and $\theta$ but with different thickness. The dimensions of each specimen were summarized in Table 1. The length and width were measured by optical microscope, and the thickness was evaluated using a laser displacement sensor LT-9500 and LT-9010M (Keyence) at a temperature where the shape of the ribbons became flat.

Measurements

The ribbon specimens were immersed in the temperature controllable bath of silicone oil. In the oil bath, the specimens were subjected to no mechanical constraint. No swelling of the specimens occurred because silicone oil was non-solvent. The shape of the ribbon specimens was observed with an optical microscope with CCD camera. The spiral pitch and diameter of the ribbons were evaluated as a function of temperature. The temperature was varied stepwise after confirming the equilibration of the shape of the ribbon at each temperature.

**Results and Discussion**

Figures 4a and 4b show the micrographs of TNE-X-1 and TNE-X-2 at various temperatures ($T$), respectively. Both specimens form a shape of spiral ribbon. The structural parameters such as spiral pitch ($p_s$) and diameter ($d$) are strongly $T$-dependent including the reversal of the handedness; the $T$ dependence of the inverse of each structural parameter for TNE-X-1 and TNE-X-2 is illustrated in Figure 5(a-b), respectively. In these figures, for convenience, the parameter values for the left- and right-handed spiral ribbons are defined to be positive and negative, respectively, and the condition of $p_s^{-1} = d^{-1} = 0$ corresponds to the flat shape. The temperature in the figures ($\tau$) is reduced using the nematic-isotropic transition temperature ($T_{NI}$ = 368 K) and the temperature where the ribbons become flat in shape ($T_{flat}$ = 353 K): $\tau \equiv (T - T_{flat})/(T_{NI} - T_{flat})$; $\tau = 0$ at $T = T_{flat}$ and $\tau = 1$ at $T = T_{NI}$.

In the high-temperature isotropic state of $T > T_{NI}$, a right-handed spiral ribbon is formed, and the structural parameters are $T$-independent. In the nematic state of $T < T_{NI}$, $p_s$ and $d$ increases as $T$ decreases, and the shape becomes flat at a temperature around 353 K. Further cooling results in the chirality reversal of the spiral shape, i.e. the formation of left-handed spiral ribbons. No appreciable difference in the structural parameters between heating and cooling processes across $T_{NI}$ was observed, which shows that the twist configuration of the mesogens is effectively imprinted and memorized in the elastomer matrix even after the removal of the chiral dopant employed in the stage of cross-linking.

The markedly $T$-dependent shape in the nematic state of $T < T_{NI}$ indicates that the shape variation is caused by a change in local nematic order driven by $T$ variation. The temperature where the shape of the ribbons becomes flat ($T_{flat} \approx 353$ K) is different from the "original flat temperature" at which the cross-linking reaction was conducted ($T_0 \approx 318$ K). This is because a considerable degree of spiral deformation was accompanied by a large volume reduction of ca. 50% from the gel state after cross-linking. TNE-X-1 shows larger values of $|p_s^{-1}|$ and $|d^{-1}|$ than TNE-X-2, while the two specimens are similar in $T_{\text{flat}}$. This shows that as the thickness is smaller, the films form more tightly spiral ribbons.

In the cases of on-axis L- and S-geometry, the TNE films form the helicoids twisting around the central line when the film's width/thickness ratio is smaller than a critical value (ca. 8.5), which was demonstrated in our previous work[11]. Importantly, both TNE films with off-axis X-geometry form spiral shape in the whole temperature range (Fig. 4), although they have the smaller width/thickness ratios (3.1 and 1.4) than the critical value. This finding clearly indicates that not only sufficiently small width/thickness ratio but also on-axis L- or S- geometry are required for the formation of helicoids. This will be further discussed with the simulation results for the geometry ($\theta$) effects on shape in the later sections.

**Simulation approach**

We model the temperature-driven shape deformation in twist nematic elastomers using a 3-d, nonlinear finite element elastodynamics algorithm [20]. The LCE sample is discretized into an unstructured, three dimensional tetrahedral mesh using open source software [21]. The sample's microstructure is described by the director field $\vec{n}$ which is defined within each tetrahedron. The nematic director field is thus represented as a piece-wise constant. The Hamiltonian of the system is [22]:

$$H = \sum_p \frac{1}{2} m_p v_p^2 + \sum_t \frac{1}{2} V^t C_{ijkl} \varepsilon_{ij}^t \varepsilon_{kl}^t - \alpha \sum_t V^t \varepsilon_{ij}^t \left( Q_{ij} - Q_{ij}^{flat} \right)^t \qquad (1)$$

The first term is the system kinetic energy calculated as a sum over the nodes $p$ in the mesh; the instantaneous node velocity is $v_p$, and the node mass $m_p$ is determined via the lumped mass approximation [23]. The second term describes the elastic potential energy of an isotropic solid as a sum over tetrahedral elements $t$. The stiffness tensor $C_{ijkl}$ depends on the bulk and shear moduli of the isotropic solid [24], and $\varepsilon_{ij}$ is the Green-Lagrange nonlinear strain tensor [25]. By using this nonlinear form of the strain tensor, we ensure that the elastic potential energy is invariant under rotation of the sample. The third term describes the coupling between nematic order and mechanical strain [26]. Higher order couplings are allowed, but this form is sufficient to capture the essential features of the strain-order coupling. Due to strong cross-linking in our nematic polymer samples, the director rotates with the body frame but does not otherwise evolve in response to strain [4][8]. Consequently, the order parameter tensor $Q_{ij} = S(3n_i n_j - \delta_{ij})/2$ depends on the configuration of $\vec{n}$ (which is fixed in the body frame) and the scalar nematic order parameter $S$ which varies with temperature.

The liquid crystal elastomer sample's shape evolution is characterized by the motion of nodes in the mesh, which follows with effective forces $\vec{F}_p = m_p \vec{a}_p$ calculated as the derivative of the potential energy with respect to node displacements. Equations of motion are integrated numerically using the velocity Verlet algorithm [27].

To represent temperature change in the FEM simulation, we adjust the magnitude of the scalar order parameter according to: $T/T_{NI} = 1.01 - (\alpha S/3.3\mu)^{3/2}$ where $\mu = C_{xyxy}$ is the material shear modulus [24]. This relation was obtained by fitting experimental, numerical and analytical data for the thermally induced elongation of a NE ribbon with planar anchoring inducing uniform $\vec{n}$ (without twist) oriented along the ribbon's long axis [11] and the same cross-link concentration as the TNE specimens. In our FEM simulation, we assume that the undistorted state occurs at $T_{flat}/T_{NI} = 0.324$, with $\alpha S_{flat}/\mu = 2.566$. After applying a change in temperature, we integrate our finite element simulation forward in time to observe shape evolution. In the absence of damping forces, the sum of kinetic and potential energy is conserved to high precision. To relax the sample toward its new elastic equilibrium, we introduce dissipation in the form of a drag force applied to each node proportional to its momentum. This dissipation removes kinetic energy from the system and allows the sample to relax to a minimum of its total potential energy. While metastable elastic minima are common in chiral elastic strips [28] we adjusted the strength of the dissipation in our model to allow each sample to relax to its global minimum energy configuration.

We simulated two sample sizes with aspect ratio (length-width-thickness) of 500-34-10 and 500-13-10, which match the width-to-thickness ratio of experimental samples TNE-X-1 and TNE-X-2. Twisted director microstructure of these samples with X-geometry is shown in Figure 2. We start the sample at temperature $T_{flat}$ and heat gradually to a target temperature over 125,000 time steps, then hold $T$ fixed until the sample relaxes to elastic equilibrium. The pitch and diameter of the spiral structures arising from the FEM simulation are determined by fitting a helical curve to the node positions along one edge of the ribbon [29].

**Shape and Chirality Dependence on Temperature**

Simulation results for both TNE-X-1 and TNE-X-2 samples with small offset angle $\theta = -5°$ are shown in Fig. 6, showing the equilibrium shape at various temperatures. In all cases we observe spiral geometries in agreement with the experimental observations in Fig. 4. This result is surprising because for the special case $\theta = \pm 45°$, which corresponds to L- or S- geometry, such narrow samples are always helicoid in shape [11]. Our simulations thus confirm the experimental finding that a change of offset angle for the twisted director microstructure strongly affects shape selection.

In Figure 5(a-b) we compare our simulation results with experiment for the $T$-dependence of spiral pitch and diameter in samples TNE-X-1 and TNE-X-2. We performed simulations with different values of the offset angle $\theta$ and found that best agreement between simulation and experiment was $\theta = -7°$, which is slightly larger but close to the experimental value ($\theta = -5° \pm 1°$). Simulation studies show that the $T$-dependence of the spiral pitch is highly sensitive to small errors in the offset angle ($\delta\theta \sim 2°$); in contrast, spiral diameter $d$ is completely insensitive to $\delta\theta$. Figure 7 show a comparison of the $T$-dependence of spiral pitch and diameter for samples with $\theta$ ranging from -3° to -9°.

While simulation results for pitch vs. temperature are in close quantitative agreement with experiments for the high $T$-range, they predict a somewhat smaller pitch than observed experimental values in the low $T$-range. This disagreement most likely arises from slight inaccuracy of the empirical relation between the scalar order parameter and temperature, due to variability of cross-link density among different samples.

Also in Figure 6(a-b) we observe that our simulation shows a transition of macroscopic chiral sense as a function of temperature. For $T > T_{flat}$ ($\tau > 0$), we find right-handed spirals, and for $T < T_{flat}$ ($\tau < 0$) we find left-handed spirals; the transition occurs at $\tau = 0$. We did not find any dependence of macro-chirality on the sample's aspect ratio.

**Shape and chirality dependence on angular offset**

As discussed above, narrow elastomer ribbons with the off-axis X-geometry always form spiral shapes, while on-axis L- and S- specimens with the same small aspect ratio form helicoidal shapes when actuated, as shown in Figure 8. This result evidently demonstrates that the L- or S-geometry – where the director at the mid-plane position is parallel to the long or short axis – is required for the formation of helicoids. The central line in helicoids is straight, which requires the local director at the mid-plane to be symmetric as in the L- and S-geometries ($\theta = +45°$ or $-45°$). Evidently, the mid-plane director in the X-geometry ($\theta \neq \pm 45°$) is not symmetric, as a result such TNE-X films form spiral ribbons independently of their aspect ratio. The spiral ribbon of TNE-X-1 in the isotropic state is right-handed in accordance with the handedness of the helicoid of TNE-S-1.

In Figure 9 we show numerical simulation results to explore the dependence of shape selection on the offset angle $\theta$. We find that there is a smooth transition from spiral ribbons to helicoids for sufficiently narrow films ($w < w_c$ where $w_c$ is the critical width for shape transition) when $\theta$ approaches from $-5°$ to $-45°$ at a constant temperature. The spiral pitch increases with $\theta$ while diameter decreases, and some helicoidal twist is observed in the spiral as $\theta$ increases. Interestingly, the perfect helicoid is formed only when $\theta$ is very close to $-45°$. This result demonstrates that the symmetry in the local director configuration at the sample mid-plane is a required condition for the formation of helicoids.

The angular offset $\theta$ not only determines the elastomer's shape, but also the elastomer's macroscopic chirality at a fixed $T$. Simulations at $\tau = 0.506$ are presented in Figure 10. These results demonstrate that macroscopic chirality switches from right-handed to left-handed as a function of $\theta$, which indicates the existence of a threshold angle $\theta_C$ for chirality transition. At this threshold, the TNE sample forms an achiral macrostructure independent of its microscopic chirality. Surprisingly, $\theta_C$ is not zero; simulations with $\theta = 0$ at $\tau = 0.506$ predict a right-handed spiral. This macroscopic chirality arises from the handedness of the twist in the director field.

Our studies determined the critical angle for chirality reversal at $\theta_C \approx +2.5°$ when $\tau = 0.506$. For these parameters, samples form macroscopic achiral shapes in the form of an open ring, even though their microscopic chirality is fixed as left-handed. The presence of an achiral shape in this chiral system shows that macroscopic chirality in TNE arises from two sources: (1) the handedness of director twist and (2) the angular offset $\theta$, which can cooperate or compete depending on the value of $\theta$.

In the high-$T$ range, the left-handed director twist induces right-handed macroscopic chirality while the angular offset can induce either left-handed ($\theta > 0$) or right-handed macroscopic chirality ($\theta < 0$). Due to the competition between these chirality sources, the sample can form achiral, left-handed or right-handed ribbons. Our results suggest a very weak right-handed influence from the director twist, an angular offset as small as +2.5° is enough to cancel it. Thus this sample forms two different achiral states at two different temperatures: a flat state with no curvature at $\tau = 0$, and a curved open ring with zero pitch at $\tau = 0.506$. The experimental characterization of $\theta_c$ at a fixed $\tau$ will be a subject in our future studies.

In summary, these simulation studies of TNE ribbons in X-, S-, and L-geometries demonstrate that off-axis angle θ strongly affects the temperature-induced shape and macro-chirality transitions in TNE elastomers. Our simulations predict that TNE-X samples with small values of θ form spiral ribbons instead of helicoids, in close agreement with experimental observations. Simulations also show that samples with a larger angular offset can form a family of more complex chiral shapes with a continuous variation between spiral for $\theta = -5°$ and helicoids for $\theta = -45°$, as shown in Fig. 9. The macroscopic chirality of TNE samples at a finite temperature also depends on $\theta$ as shown in Fig. 10.

## Conclusions

We have performed both experimental and simulation studies of off-axis twist nematic elastomer ribbons, and demonstrated that both shape selection and chirality transitions depend sensitively on the value of the offset angle $\theta$. For small offset angles, samples form spirals. As $\theta$ increases we observe increasing helicoidal twist coexisting with the spiral deformation; and for the special case $\theta=45°$, purely helicoidal ribbons arise. The formation of helicoids in TNE reported in previous studies requires special symmetry in the sample, with the local director in the sample mid-plane lying parallel to either the ribbons long axis (L-geometry, $\theta = +45°$) or short axis (S-geometry, $\theta = -45°$). Both experimental and simulation data demonstrate that sample pitch and diameter vary with temperature

Both experiments and numerical simulations also show transitions in the macroscopic chiral sense of TNE ribbons as a function of temperature. This feature of TNE materials occurs for all director geometries studied (S-, L- and X-geometries). In the high-$T$ range samples form chiral structures with opposite handedness to the structures observed in the low-$T$ range with a transition at $T=T_{flat}$ (reduced temperature $\tau = 0$).

Simulations show that the macroscopic chirality in TNE arises due to the combination of the handedness of director twist and the offset angle $\theta$. These sources can either compete or cooperate to produce achiral, right-handed or left-handed shapes. Achiral shapes (open rings) are observed at a critical angle $\theta_C$ where the two effects cancel each other.

**Table 1.  Sample Characteristics**

|  | $\theta$ (º) | Width (μm) | Thickness (μm) | Length (mm) |
|---|---|---|---|---|
| TNE-X-1 | −5 | 110 | 35.2 | 9.3 |
| TNE-X-2 | −5 | 120 | 88.0 | 9.4 |
| TNE-L-1 [a] | +45 | 230 | 35.2 | 9.2 |
| TNE-S-1 [a] | −45 | 220 | 35.2 | 9.8 |

[a] The data was reproduced from reference [11].

## Acknowledgments

This work was supported in part by the Grant-in-Aid on Priority Area "Soft Matter Physics" (No.21015014) and that for Scientific Research (B) (No. 16750186) from the Ministry of Education, Culture, Sports, Science, and Technology (MEXT) of Japan (K.U.) and in part by NSF-DMR 1106014, the Wright Center of Innovation for Advanced Data Management and Analysis, and by the Ohio Board of Regents.

**References**

[1] M. Warner and E. M. Terentjev, Liquid Crystal Elastomers (International Series of Monographs on Physics). Oxford University Press, USA, 2003.

[2] R. Z. C. Ohm, M. Brehmer, and R. Zentel, "Applications of Liquid Crystalline Elastomers," in Liquid Crystal Elastomers: Materials and Applications, vol. 250, W. H. de Jeu, Ed. Berlin, Heidelberg: Springer Berlin Heidelberg, 2012.

[3] J. Naciri, A. Srinivasan, H. Jeon, N. Nikolov, P. Keller, and B. R. Ratna, "Nematic Elastomer Fiber Actuator," Macromolecules, vol. 36, no. 22, pp. 8499–8505, Nov. 2003.

[4] C. D. Modes and M. Warner, "Blueprinting nematic glass: Systematically constructing and combining active points of curvature for emergent morphology," Physical Review E, vol. 84, no. 2, Aug. 2011.

[5] C. L. van Oosten, K. D. Harris, C. W. M. Bastiaansen, and D. J. Broer, "Glassy photomechanical liquid-crystal network actuators for microscale devices.," The European physical journal. E, Soft matter, vol. 23, no. 3, pp. 329–36, Jul. 2007.

[6] G. N. Mol, K. D. Harris, C. W. M. Bastiaansen, and D. J. Broer, "Thermo-Mechanical Responses of Liquid-Crystal Networks with a Splayed Molecular Organization," Advanced Functional Materials, vol. 15, no. 7, pp. 1155–1159, Jul. 2005.

[7] K. M. Lee, T. J. Bunning, and T. J. White, "Autonomous, hands-free shape memory in glassy, liquid crystalline polymer networks.," Advanced materials (Deerfield Beach, Fla.), vol. 24, no. 21, pp. 2839–43, Jun. 2012.

[8] C. D. Modes, M. Warner, C. Sánchez-Somolinos, L. T. de Haan, and D. Broer, "Mechanical frustration and spontaneous polygonal folding in active nematic sheets," Physical Review E, vol. 86, no. 6, p. 060701, Dec. 2012.

[9] M. Yamada, M. Kondo, J. Mamiya, Y. Yu, M. Kinoshita, C. J. Barrett, and T. Ikeda, "Photomobile polymer materials: towards light-driven plastic motors.," Angewandte Chemie (International ed. in English), vol. 47, no. 27, pp. 4986–8, Jan. 2008.

[10] Y. Geng, P. L. Almeida, S. N. Fernandes, C. Cheng, P. Palffy-Muhoray, and M. H. Godinho, "A cellulose liquid crystal motor: a steam engine of the second kind.," Scientific reports, vol. 3, p. 1028, Jan. 2013.

[11] Y. Sawa, F. Ye, K. Urayama, T. Takigawa, V. Gimenez-Pinto, R. L. B. Selinger, and J. V Selinger, "Shape selection of twist-nematic-elastomer ribbons.," Proceedings of the National Academy of Sciences of the United States of America, vol. 108, no. 16, pp. 6364–8, Apr. 2011.

[12] R. Oda, I. Huc, M. Schmutz, S. J. Candau, and F. C. MacKintosh, "Tuning bilayer twist using chiral counterions.," Nature, vol. 399, no. 6736, pp. 566–9, Jun. 1999.

[13] J. V. Selinger, M. S. Spector, and J. M. Schnur, "Theory of Self-Assembled Tubules and Helical Ribbons," The Journal of Physical Chemistry B, vol. 105, no. 30, pp. 7157–7169, Aug. 2001.

[14] J. Adamcik and R. Mezzenga, "Proteins Fibrils from a Polymer Physics Perspective," Macromolecules, vol. 45, no. 3, pp. 1137–1150, Feb. 2012.

[15] R. L. B. Selinger, J. V Selinger, A. P. Malanoski, and J. M. Schnur, "Visualizing chiral self-assembly.," Chaos (Woodbury, N.Y.), vol. 14, no. 4, p. S3, Dec. 2004.

[16] R. L. B. Selinger, J. V Selinger, A. P. Malanoski, and J. M. Schnur, "Shape selection in chiral self-assembly.," Physical Review Letters, vol. 93, no. 15, p. 158103, Oct. 2004.

[17] Z. Chen, C. Majidi, D. J. Srolovitz, and M. Haataja, "Continuum Elasticity Theory Approach for Spontaneous Bending and Twisting of Ribbons Induced by Mechanical Anisotropy," http://arxiv.org/abs/1209.3321, Sep. 2012.

[18] L. Teresi and V. Varano, "Modeling helicoid to spiral-ribbon transitions of twist-nematic elastomers," Soft Matter, 2013, DOI: 10.1039/C3SM27491H.

[19] Y. Sawa, K. Urayama, T. Takigawa, A. DeSimone, and L. Teresi, "Thermally Driven Giant Bending of Liquid Crystal Elastomer Films with Hybrid Alignment," Macromolecules, vol. 43, no. 9, pp. 4362–4369, May 2010.

[20] B. L. Mbanga, "Hybrid Particle-Finite Element Elastodynamics Simulations of Nematic Liquid Crystal Elastomers," PhD Dissertation, Kent State University, 2012.

[21] "SALOME Platform." [Online]. Available: http://www.salome-platform.org/. [Accessed: 02-May-2012].

[22] B. Mbanga, F. Ye, J. Selinger, and R. Selinger, "Modeling elastic instabilities in nematic elastomers," Physical Review E, vol. 82, no. 5, Nov. 2010.

[23] R. Rudd and J. Broughton, "Coarse-grained molecular dynamics and the atomic limit of finite elements," Physical Review B, vol. 58, no. 10, pp. R5893–R5896, Sep. 1998.

[24] M. P. Marder, Condensed Matter Physics. Wiley-Interscience, 2000.

[25] M. Warner and E. M. Terentjev, Liquid Crystal Elastomers (International Series of Monographs on Physics). Oxford University Press, USA, 2003, p. 424.

[26] P. G. de Gennes and J. Prost, The Physics of Liquid Crystals (International Series of Monographs on Physics). Oxford University Press, USA, 1995.

[27] D. C. Rapaport, The Art of Molecular Dynamics Simulation. Cambridge University Press, 2013.

[28] Z. Chen, Q. Guo, C. Majidi, W. Chen, D. J. Srolovitz, and M. P. Haataja, "Nonlinear Geometric Effects in Mechanical Bistable Morphing Structures," Physical Review Letters, vol. 109, no. 11, p. 114302, Sep. 2012.

[29] P. C. Kahn, "Defining the axis of a helix," Computers & Chemistry, vol. 13, no. 3, pp. 185–189, Jan. 1989.

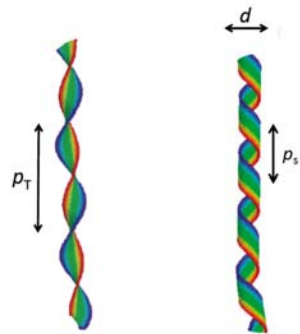

**Figure 1** (a) Helicoid shape displays negative Gaussian curvature with chiral pitch $p_T$ and a straight central line. (b) Spiral ribbon has zero Gaussian curvature and a curved central line with chiral pitch $p_S$ and diameter $d$ (From reference [14].)

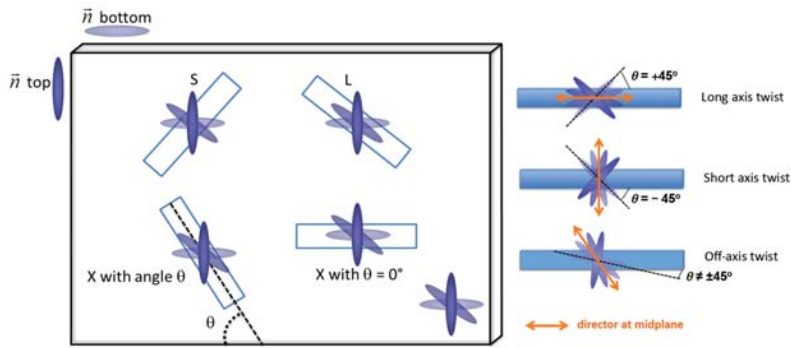

**Figure 2** Twist nematic elastomer ribbons with three types of geometry for director configuration.

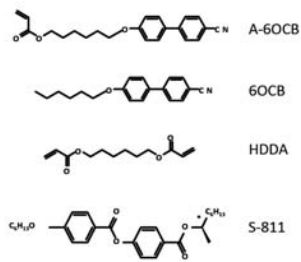

**Figure 3** Chemical structures of employed compounds

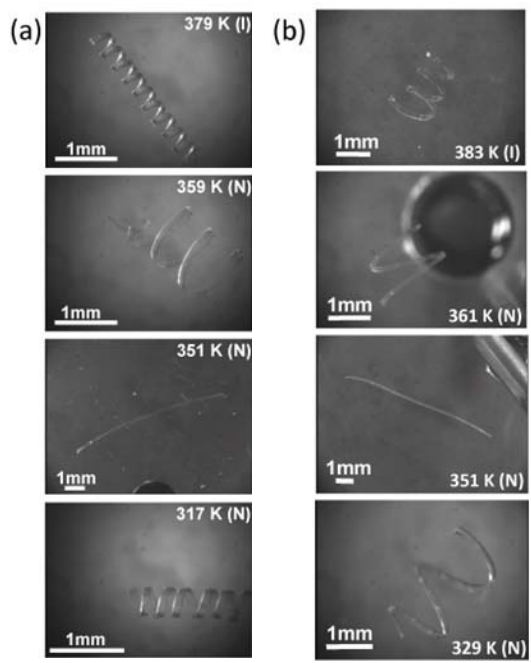

**Figure 4** The equilibrium shapes of (a) TNE-X-1 and (b) TNE-X-2 with the same width but different thickness as a function of temperature. I and N denote the isotropic and nematic states, respectively. Both TNE films form spiral ribbons, and the spiral handedness is right and left in the $T$ region of $T > T_{flat}$ and $T < T_{flat}$, respectively, where $T_{flat}$ is ca. 353 K. The thinner elastomer (a) forms more tightly spiral ribbon than (b).

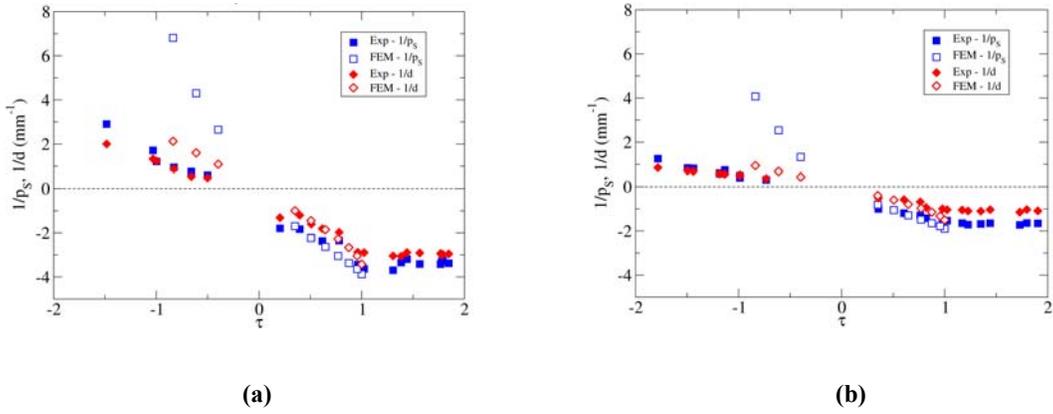

**(a)** **(b)**

**Figure 5** The inverses of spiral pitch $p_s$ and diameter $d$ of (a) TNE-X-1 and (b) TNE-X-2 as a function of reduced temperature $\tau = (T-T_{flat})/(T_{NI}-T_{flat})$, comparing FEM simulation results with experiment. Temperature is reduced by nematic-isotropic transition temperature ($T_{NI}$ = 368 K) and the temperature where the ribbons become flat in shape ($T_{flat}$ = 353 K). The pitches of the left- and right-handed twist are defined to be positive and negative, respectively.

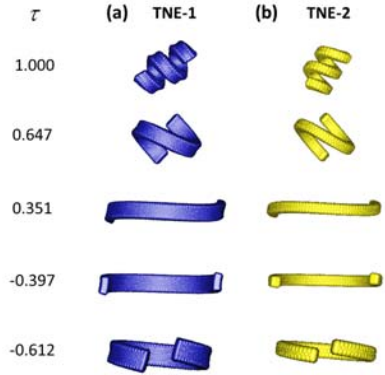

**Figure 6** Simulation results showing temperature dependence of TNE with X geometry; samples have the same width/thickness ratio and offset angle $\theta$ as (a) TNE-X-1 and (b) TNE-X-2. Both samples switch their chiral sense at $T=T_{flat}$ which corresponds to $\tau=0$.

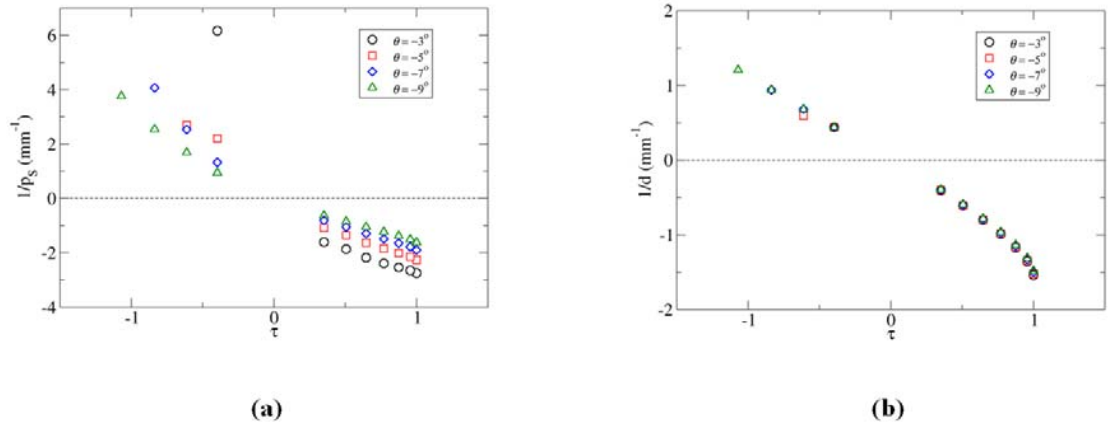

**Figure 7** Simulation results for the temperature dependence of (a) inverse spiral pitch $1/p_s$ and (b) inverse diameter $1/d$ of the TNE ribbons with various $\theta$. The width/thickness ratio matches the experimental sample TNE-X-2. Pitch shows strong dependence on offset angle but diameter is nearly invariant. The pitches of the left- and right-handed twist are defined to be positive and negative, respectively.

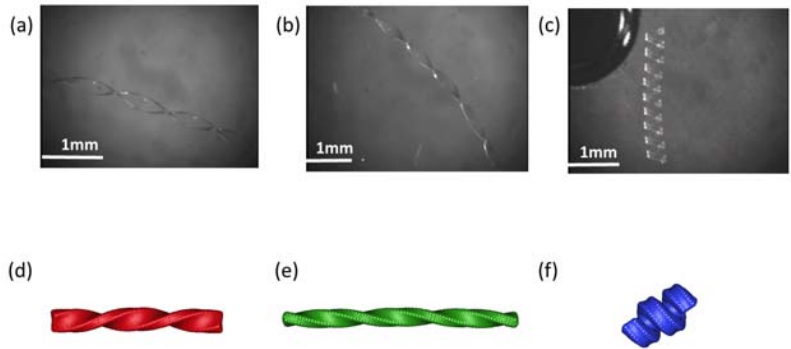

**Figure 8** Micrographs of (a) TNE-L-1, (b) TNE-S-1 and (c) TNE-X-1 at 384 K. TNE-L-1 and TNE-S-1 form left- and right-handed helicoids, respectively. TNE-X-1 has a width well below the critical width for shape transition $w_c$, but it forms a right-handed spiral. Simulations of ribbons with (d) L-geometry, (e) S-geometry and (f) X-geometry with $\theta = -5°$ at $\tau = 1.000$ show the same result. Sample aspect ratio in simulation images corresponds with TNE-X-1.

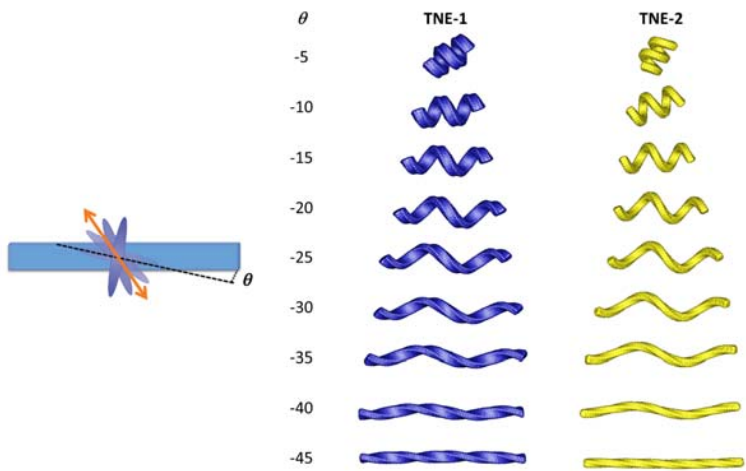

**Figure 9** Simulation results for the shapes of TNE ribbons with the same width/thickness ratio as TNE-X-1 and TNE-X-2 as a function of $\theta$ at $\tau = 1.000$

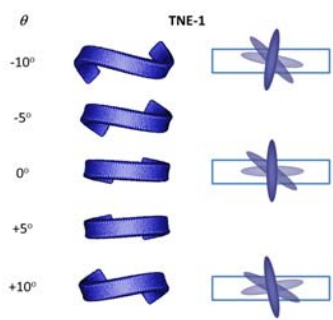

**Figure 10** Simulation results for the shapes of TNE ribbons with the same width/thickness ratio as TNE-X-1 as a function of $\theta$ at reduced temperature $\tau = 0.506$. When angular offset is zero or negative we observe the formation of left-handed ribbons, while positive angles shown here produce right-handed shapes.